# Stabilizing effect of nuclear quadrupole interaction on the polarization of electron-nuclear spin system in a quantum dot

R.I. Dzhioev, V.L. Korenev[*]

*A. F. Ioffe Physical Technical Institute, St. Petersburg, 194021 Russia*

*Nuclear quadrupole interaction eliminates the restrictions imposed by hyperfine interaction on the spin coherence of an electron and nuclei in a quantum dot. The strain-induced nuclear quadrupole interaction suppresses the nuclear spin flip and makes possible the zero-field dynamic nuclear polarization in self-organized InP/InGaP quantum dots. The direction of the effective nuclear magnetic field is fixed in space, thus quenching the magnetic depolarization of the electron spin in the quantum dot. The quadrupole interaction suppresses the zero-field electron spin decoherence also for the case of non-polarized nuclei. These results provide a new vision of the role of the nuclear quadrupole interaction in nanostructures: it elongates the spin memory of the electron-nuclear system.*

PACS: 78.67.–n, 73.21.–b, 85.75.–d

# INTRODUCTION

An electron in a semiconductor quantum dot (QD) interacts with a large number $N \sim 10^5$ of the surrounding lattice nuclei. If one considers the electron spin as a solid-state quantum bit the spin decoherence should be overwhelmed. Hyperfine interaction (with hyperfine constant $A \approx 100\,\mu eV$) imposes strict limits on the electron spin coherence time [1] at zero magnetic field: the initial dephasing in a random quasi-static nuclear field within a few nanoseconds [2] is followed by the decoherence over the microsecond timescale, because the nuclear field changes in time due to the precession of nuclear spins in the effective electron field (Knight field) of hyperfine interaction. An external magnetic field effectively decouples the electron and the nuclei, suppressing transitions with mutual electron and nuclear spin flips. Such a "brute force" technique extends the longitudinal electron spin relaxation time up to milliseconds [3]. More refined approach involves the optical cooling of nuclei down to ultra low temperature $\theta \sim 10^{-6}\,K$ [4]: a relatively weak time-averaged Knight field induces a substantial nuclear polarization. Polarized nuclei create an effective magnetic field due to the hyperfine interaction that suppresses the electron spin relaxation similarly to the external field. The dynamic nuclear polarization in a low external field (*1-10 G*) was observed in charged InP/InGaP [5], InGaAs/GaAs [6], InAs/GaAs [7] and CdSe/ZnSe [8] QDs.

Here we propose another approach and show that the nuclear quadrupole interaction eliminates the restrictions imposed by hyperfine interaction on the spin coherence of the electron and nuclei in a quantum dot. The strain-induced quadrupole interaction suppresses the spin flip of nuclei and makes possible the zero-field dynamic nuclear polarization in quantum dots. The direction of the effective magnetic field of polarized nuclei is fixed in space, thus quenching the magnetic depolarization of the electron spin in the quantum dot. The quadrupole interaction suppresses the zero-field electron spin decoherence also for the case of non-polarized nuclei.

Therefore the nuclear quadrupole interaction elongates the spin lifetime of the electron-nuclear system in quantum dots.

## MEASUREMENT TECHNIQUE

The structures were grown by MOVPE on [001]-GaAs substrate and contained nano-sized InP islands (120 nm in diameter with density $3 \cdot 10^9 \, cm^{-2}$) restricted by InGaP barrier. The layers contained donor impurities $\approx 10^{15} \, cm^{-3}$. The samples were placed in a liquid-helium cryostat (T=4.2 K) and pumped by a 10 W/cm$^2$ He-Ne laser (1.96 eV) above wetting layer. The photoluminescence (PL) circular polarization $\rho_c$ was measured in the reflection geometry within the maximum of PL intensity. The external magnetic field $\vec{B}$ was applied in Voigt or in Faraday geometries. Measurements both in the presence and in the absence of dynamic nuclear polarization were carried out. To avoid the dynamic polarization of nuclei the helicity of light was alternated in sign at a frequency of 26.61 kHz with a photoelastic quartz modulator in the excitation channel and a fixed quarter-wave plate as analyzer. In this case there is not enough time for the nuclear spin to follow the polarization of electrons [9]. To initiate the Overhauser effect the polarizer and analyzer were interchanged.

Excitation of QD nanostructures by circularly polarized light provides the optical pumping of resident QD electrons [10, 11, 12]. A plausible scenario is the capture of single carriers into the dots. For example, the hole is captured first, forming a neutral exciton (an optically allowed, bright exciton with angular momentum projection $\pm 1$ or an optically forbidden, dark exciton with projection $\pm 2$ onto [001]). This is followed by the trapping of the photo-electron. If the exciton recombines first, then the optically oriented electron becomes a resident. Thus the optical pumping takes place by the replacement of non-polarized electrons by the oriented ones, similar to the usual optical orientation in bulk n-type semiconductors [4]. Optical orientation of both neutral excitons and resident electrons can be detected by the degree $\rho_c$ of the PL circular polarization of the QD ensemble. Various transitions related to the neutral or charged excitons contribute to the PL polarization. The polarized PL of empty (of electrons) islands comes from the recombination of neutral bright excitons $X^0$. PL of singly charged islands originates from the ground state of $X^-$ trion – the complex of two electrons with antiparallel spins and one hole. In turn, polarization of trions is determined by the polarization of neutral excitons as well as the mean electron spin at the moment of trion formation [10]. Usually the helicity of $X^0$ emission coincides with the helicity of pumping light. In contrast with it the helicity of $X^-$ emission is often opposite to the pump polarization (so-called negative polarization) [10, 11, 12].

The Hanle effect measurements (Voigt geometry) separate the contributions of neutral excitons and resident electrons to the circular polarization of the PL. In a low magnetic field (*B~100 G*) only resident electron spins undergo Larmor precession [13] thus depolarizing PL. In contrast with the Hanle effect, the measurements in Faraday geometry reveal the fine structure of the neutral excitons forming trions. The electron-hole anisotropic exchange interaction [14] mixes +1 and -1 bright states (it comes from the QD anisotropy and induces splitting by $\delta_b = 10 - 100\,\mu eV$) as well as dark states +2 and –2 (split by $\delta_d = 1 - 2\,\mu eV$). As a result the depolarization of neutral excitons takes place. An external magnetic field in Faraday geometry eliminates the mixing, thus restoring their optical orientation.

## RESULTS

Figure 1 shows the dependence $\rho_c(B)$ in Voigt ($\vec{B} \perp [001]$, Fig.1a) and Faraday ($\vec{B} \parallel [001]$, Fig.1b) geometries. Polarization is negative which is a striking signature of $X^-$ trions. The Hanle effect of resident electrons (filled circles on Fig.1a) is measured in the absence of the dynamical nuclear polarization (excitation by light of alternating helicity). The Hanle curve can be fitted by Lorentzian with the halfwidth $B_{1/2} \approx 50$ G. Taking into account the electron g-factor g=1.6 [15] one can estimate the electron spin lifetime according to formula $T_s = \hbar/\mu_B g B_{1/2} \approx 1.4$ ns [4, 16]. One can see that a complete PL depolarization takes place. This means that the neutral excitons forming trion are non-polarized. An external magnetic field in Faraday geometry restores the optical orientation of excitons. Because of the inequality $\delta_d \ll \delta_b$ one restores the orientation of dark excitons in low fields enhancing the circular polarization of PL (filled circles on Fig.1b) followed by the restoration of orientation of bright states in larger (~10 kG) magnetic field (this range is not shown, see Ref. [10] for details). The dependences $\rho_c(B)$ are symmetrical with respect to the change of $B$ sign.

The presence of dynamical nuclear polarization (open circles on Figs.1a, b are measured under excitation by light of constant helicity) changes situation substantially. The curve of restoration of the trion optical orientation is shifted (Fig. 1b) by the nuclear magnetic field value about 80 G (the curve is broadened due to the non-uniform nuclear field distribution). Moreover, the zero-field polarization $\rho_c(B=0)$ value is also affected [5]. It means that the dynamic nuclear polarization persists at zero field, too. For InAs QDs this effect was interpreted [7] as the Knight field-enabled nuclear polarization. In our case (at least), it is the result of electric quadrupole interaction suppressing the depolarization of In nuclei by local magnetic fields $B_L \sim 1$ G of the surrounding nuclei. The Hanle effect measurements (open circles on Fig.1a) support this conclusion. The fact is that the Hanle curve (it is symmetrical with respect to the ordinate axis) halfwidth is 150 G. This is *three times larger* than the value $B_{1/2} = 50\,G$ in the absence (filled circles on Fig.1a) of nuclear polarization [17]. This result contradicts the standard theory of optical orientation of electrons and nuclei in the non-strained GaAs-type semiconductor (i.e. without quadrupole effects). Indeed the nuclear field enhances the depolarizing effect of the external field on the spin of electrons [9] leading to the *narrowing* of the Hanle curve in striking contrast with the result on Fig.1a (see also Fig.2a for normalized data). Thus the Hanle curve broadening cannot be explained by the Knight-field-enabled nuclear polarization that should enhance the low-field depolarization of PL.

It is important to note that the values of the nuclear-field-induced shift of the $\rho_c(B)$ in Faraday geometry (Fig.1b) and the Hanle curve halfwidth (Fig.1a) in the presence of nuclear polarization are comparable. Therefore we have to suggest the existence of the nuclear field *whose direction is fixed in space near the [001] axis and is non-zero even for B=0*. It both shifts the $\rho_c(B)$ dependence in Faraday geometry and blocks the Larmor precession of electron spin in Voigt geometry (Fig.2b). The fixing of the nuclear field direction in InP islands can be related with the quadrupole interaction of indium nuclei only [18] (95.5 % of $^{115}$In and 4.5 % of $^{113}$In, both with *I=9/2* [19]). It is essential in the low-field region when Zeeman splitting of nucleus $\gamma\hbar B$ ($\gamma$ - nuclear gyromagnetic ratio) is less than the quadrupole splitting $h\nu_Q$.

The dynamic nuclear polarization in the presence of strong quadrupole interaction (QI) was studied intensively in AlGaAs bulk solid solutions [9]. Substitution of Ga atoms by Al atoms leads to the QI of As nuclei. The main axes of the QI were directed along {111}. In our case the binary compound InP forms islands and QI is absent in zero approximation [20]. However, a very large strain (due to 3.7 % lattice mismatch between InP and InGaP) takes place in the system under study [21] with the main axis being close to the growth direction. Strain will induce the QI of indium nuclei and will dominate in a low magnetic field. In the simplest case of uniaxial strain along z-axis ($e_{zz} \neq 0$) the QI Hamiltonian $h\nu_Q(\hat{I}_z^2 - I(I+1)/3)/2$ is determined by the only constant [22]

$$\nu_Q = \frac{3eV_{zz}Q}{2I(2I-1)h}$$

where $Q = 0.76 \cdot 10^{-24}$ cm$^2$ is the quadrupole moment of $^{115}$In [19]. Strain induces the electric field gradient (EFG) $V_{zz} = S_{11}e_{zz}$, where the constant $S_{11} = 2\cdot 10^{16}$ statcoulombs/cm$^3$ was measured experimentally for the case of $^{115}$In nuclei in InP [23]. Estimating the z-component of deformation tensor as $e_{zz} = 0.02$ (2 %) we get $\nu_Q$ about 1 MHz. The constant $\nu_Q$ exceeds the precession frequency of In nuclei $\gamma B/2\pi$ =933 Hz/G [19] up to the field values $B \approx 1000\, G$.

Therefore, *QI dominates over the nuclear Zeeman interaction for the entire field range in Figs.1,2*. In zero field an axially symmetric QI lifts the degeneracy of In nuclei, grouping the energy levels in pairs $m = \pm 1/2, \pm 3/2, ..., \pm 9/2$ with the same module of momentum projection *m* onto z-axis (Fig.2b). In a weak magnetic field ($\gamma B << 2\pi \nu_Q$) the Zeeman energy is a small perturbation. Then nuclei in states $\pm m$ can be considered as quasiparticles with pseudospin ½ [9]. Zeeman interaction of these quasiparticles is characterized by strongly anisotropic g-factor with the main axes of g-tensor coinciding with those of EFG. In the first order the transverse magnetic field does not split the states $+m$ and $-m$ for $|m| > 1/2$, which leads to zero transverse g-factor components (the longitudinal component $g_{zz} = 2|m|$). As a result local magnetic fields $B_L$ of surrounding nuclei do not destroy the z-component of mean nuclear spin $\vec{I}_N$ even in zero magnetic field [24]. A well-known example [9] is the increase of spin relaxation time of As nuclei due to the magneto-dipole interaction up to $T_2' \sim T_2/g_\perp^2$ in an AlGaAs solid solution, where $T_2 \sim (\gamma B_L)^{-1} \approx 100\,\mu s$, transverse g – factor Lande for nuclear quasiparticles $g_\perp \sim 0.1$ results from a small deviation of EFG tensor from axial symmetry. Time $T_2' \sim 10\,ms$ becomes comparable with the spin-lattice relaxation time. Therefore the stabilizing effect of QI should be carefully analyzed in *any* (self-organized) quantum dot system having nuclei with large spin *I>1/2*. For example, the QI may explain qualitatively the zero-field nuclear polarization in InP [5] and InAs [7] dots as well as the surprising long-term conservation of spin polarization in the electron-nuclear spin system of InGaAs QD [6].

Under optical orientation conditions the mean spin of quadrupole-split nuclei is $\vec{I}_N \propto (\vec{S} \cdot \vec{n})\vec{n}$, with its direction being determined by the unit vector $\vec{n}$ along quadrupole axis rather than by the external magnetic field $\vec{B}$ [25]. These nuclei create fixed-in-space effective magnetic field $\vec{B}_N^Q = a(\vec{S} \cdot \vec{n})\vec{n}$ acting on the QD electron spin (below *a* is considered as a phenomenological fitting parameter). The nuclear field $\vec{B}_N^Q$ quenches the electron spin magnetic depolarization and enables one to explain threefold broadening of the Hanle curve in the presence of the nuclear polarization (Fig. 2a). Within the simplest model the steady state average electron spin $\vec{S}$ in the InP islands is governed by Bloch equation

$$\frac{\vec{S}_0 - \vec{S}}{T_s} + \frac{\mu_B g}{\hbar}\left[\vec{B} + \vec{B}_N^Q\right] \times \vec{S} = 0 \qquad (1)$$

where $\vec{S}_0 \| [001]$ is the zero-field mean spin; spin lifetime $T_s$ takes into account phenomenologically all contributions coming from spin relaxation and recombination [16]; the second term of Eq.(1) describes spin precession (Fig.2b) in the sum of external ($\vec{B} \perp [001]$) and nuclear ($\vec{B}_N^Q \| \vec{n}$) magnetic fields (below we put $\vec{n} \| [001]$). We neglected the contribution into the nuclear field from phosphorous nuclei whose spin ($I=1/2$) is nine times less than that of indium nuclei. The PL degree is $\rho_c \propto S_z = (\vec{S} \cdot \vec{n})$. Figure 2c shows the dependence $S_z(B)/S_0$ calculated from Eq.(1) in the absence ($a=0$, solid curve) and in the presence ($aS_0 = 5B_{1/2}$ – dashed curve) of the nuclear field. It is seen that the nuclear field of QI-perturbed nuclei does stabilize electron spin and broaden the Hanle curve by three times. Thus the experimental data can be explained by the nuclear quadrupole interaction within the standard approach [9] worked out for bulk semiconductors. Bloch equation (1) is applicable when the resident electron dwell time (correlation time) within the QD is short: electron spin has no time to make a turn about a random nuclear field. Such a situation can be easily realized under optical excitation, when the electron spin is renewed fast due to recombination or photo-induced spin relaxation processes.

## DISCUSSION

**Circular polarization of trion PL in the absence of Overhauser effect.**

As we noted above the X⁻ trions in the InP islands are formed via the binding of neutral excitons with single electron. In turn, the trion PL polarization is determined by the polarizations of bright excitons $P_b(B)$, dark excitons $P_d(B)$ and resident electrons $P_e(B)$ at the moment of trion formation [5]:

$$\rho_c(B) = \alpha P_d(B) + \beta P_b(B) + \gamma P_e(B) \qquad (3)$$

where the constants $\alpha, \beta, \gamma$ depend on the details of spin kinetics. At zero magnetic field excitons are non-polarized ($P_b = P_d = 0$) and the only contribution to the PL polarization comes from the optical orientation of single electron ($P_e \neq 0$). Figure 3 shows spectrum of PL intensity (solid line) and circular polarization (blue data points) at zero magnetic field and power density $P=30$ W/cm$^2$. The zero-field polarization is very sensitive to the power density: it disappears when the power goes to zero, as it should be in the standard scheme of optical pumping of electrons in n-type semiconductors [4].

According to Ref. [5] the total $\rho_c(B)$ dependence in Faraday geometry (filled circles on Fig.1b) results from the restoration of optical orientation of dark excitons $P_d(B)$ (bright excitons are non-polarized in the field range on Fig.1b, $P_b = 0$, so that the second term in Eq.(3) can be neglected). The dependence $P_e(B)$ of resident electron polarization represents another possible contribution into Eq.(3): the magnetic field in Faraday geometry eliminates the effect of quasi-static random nuclear fields, depolarizing the electron spin in zero field *three times* [1]. However, (**i**) the "1/3 rule" does not agree with experiment in these samples: the experimentally measured ratio - the ratio of magnetic field-induced signal in Voigt geometry to the saturated polarization value in Faraday geometry – strongly depends on experimental conditions. For example, it varies smoothly with the laser power density from zero (at the lowest power) up the 2/3 at the highest power densities *without* any remarkable feature near 1/3 (it is ½ for a particular dataset shown by filled circles on Fig.1); (**ii**) the characteristic widths of the $\rho_c(B)$ dependences in Voigt (50 G, filled circles on Fig.1a) and Faraday geometries (120 G, filled circles on Fig.1b) are different, thus contradicting with the isotropically distributed random nuclear field model [1].

These data suggest that the magnetic field-induced signals in Faraday and Voigt geometries have *different* origin. To obtain a consistent description we interpret the data on Fig.1 following the Ref. [5]: the dependence $\rho_c(B)$ in Faraday geometry comes mainly from the restoration of optical orientation of dark excitons, whereas the Hanle effect (Voigt geometry) is due to the spin precession of resident electron. The absence of any hint on "1/3 rule" can be explained by the short dwell time of the electron spin in the dot before its renewal: there is no time for the electron spin to make a lot of turns about the random nuclear field.

### Overhauser effect

As we have stated above the experimental data in the presence of nuclei (open circles on Fig.1a,b) indicate the existence of nuclear field *whose direction is fixed in space near the [001] axis and is non-zero even for B=0*. It both shifts the $\rho_c(B)$ dependence in Faraday geometry and blocks the Larmor precession of electron spin in Voigt geometry. This is qualitative result. The twofold difference between the 150 G-broadening of the Hanle curve (Fig.1b) and the $B_{sh}=80$ G shift of the restoration curve in Faraday geometry (Fig.1a) can be explained as follows. We estimate the nuclear field value $B_N^O = A\langle I\rangle/\mu_B g = 150\,G$ from the Hanle effect data, assuming that the nuclear polarization $\langle I \rangle$ originates mainly from the In nuclei. In turn, the minimum in Faraday geometry is due to the cancellation of Zeeman interaction of dark excitons, i.e. $\mu_B g_d B_{sh} + A\langle I\rangle \approx 0$, so that the $B_{sh}$ field is determined by the g-factor of the dark excitons $g_d = g + g_h$. Taking into account the hole g-factor $g_h = 1.4$ [26] we obtain $B_{sh} = gB_N^O/(g_h + g) \approx B_N^O/2 \approx 75\,G$ in good agreement with experimental value.

The observation of nuclear magnetic resonance (NMR) is a direct evidence of the Overhauser effect. An NMR signal at Larmor frequency of $^{31}$P has been detected optically [5]. At the same time the saturation of phosphor' NMR signal induces relatively small change of the total nuclear field, indicating that the nuclear polarization is mostly conserved. It is reasonable to attribute the conserved polarization to In nuclei. However, the NMR signal at the Larmor frequency of both In nuclei ($^{113}$In, $^{115}$In) has not been found. This can be explained by the QI that both shifts NMR frequencies and broadens resonance line of indium. Search for the NMR and nuclear quadrupole resonance of indium nuclei needs a special study.

Throughout the paper we have considered the effect of static nuclear quadrupole interaction on the orientation of steady state dynamic nuclear polarization. We have found that the static QI stabilizes the optical orientation of the electron-nuclear spin system in a quantum dot. In contrast to it, the fluctuations of QI will provide an additional nuclear spin relaxation channel with characteristic time $T_1^Q$, competing with Overhauser effect. The QI dynamics comes from the fluctuations of electric field gradients. The natural source of such fluctuations is due to phonons [27]. At low temperatures the phonon-induced nuclear quadrupole relaxation should be strongly suppressed.

**Single electron spin in a quantum dot in the presence of both hyperfine and quadrupole interactions**

Here we divert the reader's attention from the particular case of InP dots and discuss the general problem of the hyperfine interaction induced QD electron spin decoherence [1] in the presence of static QI. Assume that the electron spends a fairly long time $\tau_c$ (time $\tau_c$ should be much longer than the precession period $\hbar N/A \sim 1\mu s$ of nuclear spin in the Knight field) in a quantum dot before its renewal. In zero-field limit and for the non-polarized nuclei case, the first step of electron spin dephasing in a quasi-static random nuclear field [1] remains in the presence of QI, too. Indeed, the transverse (x,y) spin components of quadrupole split nuclei rotate about z-axis with characteristic angular frequency $2\pi\nu_Q \sim 10^7$ s$^{-1}$. This is much slower than the electron spin precession frequency $A/\hbar\sqrt{N} \sim 10^9$ s$^{-1}$ in random nuclear field with $N \sim 10^5$ [1], so that the nuclear field is still "frozen" and distributed isotropically at this time scale. However, the second step – the decoherence due to the Knight-field-induced precession of the nuclear spin (and the corresponding nuclear field evolution) with frequency $A/\hbar N \sim 10^6$ s$^{-1}$ [1] will be suppressed because the (x,y)-components of nuclear spin rotate about z-axis much faster than about the Knight field. Quantum mechanically one can say that the transverse components of the Knight field do not split in the first order $+m$ and $-m$ projections of a given nuclear spin for $|m| > 1/2$. Doublet degeneracy remains. As a result the z-component of both nuclear and electron spin is conserved much longer than 1 $\mu s$ [1].

## CONCLUSION

Thus the nuclear quadrupole interaction extends the limits imposed by hyperfine interaction on the spin coherence of the electron and nuclei in a quantum dot. The strain-induced nuclear quadrupole interaction suppresses the spin flip of In nuclei and makes possible the zero-field dynamic nuclear polarization in self-organized InP/InGaP quantum dots. The direction of the effective nuclear magnetic field is fixed in space, thus quenching the magnetic depolarization of the electron spin in the quantum dot. The quadrupole interaction suppresses the zero-field electron spin decoherence also for the case of non-polarized nuclei. These results provide a new vision of the role of the nuclear quadrupole interaction in nanostructures: it elongates the spin memory of the electron-nuclear system.

We are grateful to I.A. Merkulov for valuable discussions. This work was supported by RFBR 05-02-17796, Program of RAS.

R.I. Dzhioev et al, Figure 1

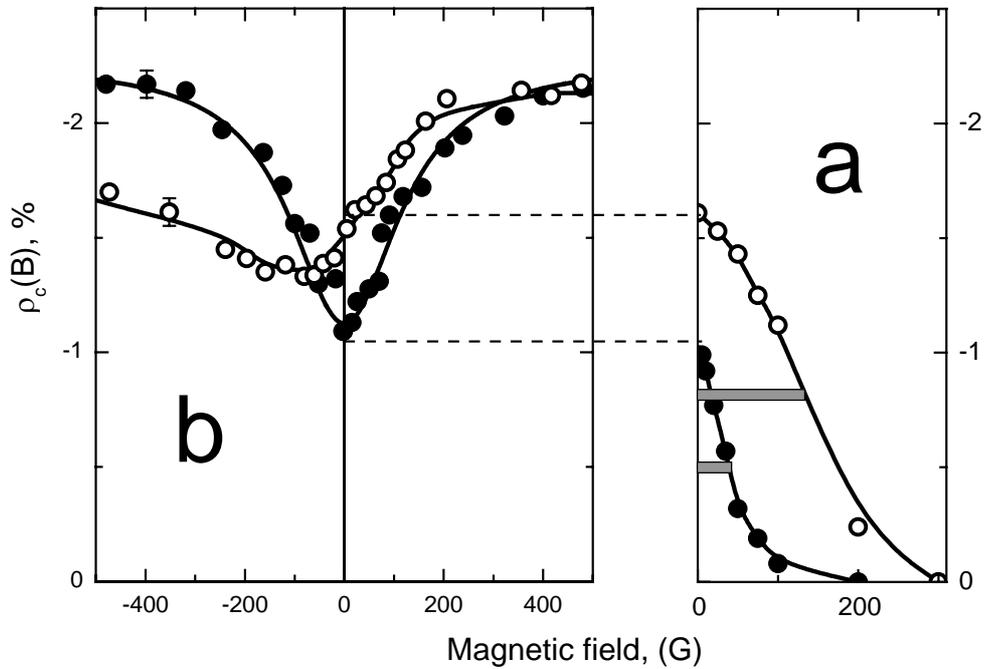

The dependence of circular polarization degree $\rho_c$ on external magnetic field *B* in Voigt (a) and Faraday (b) geometries. Filled circles are measured in the absence of nuclear polarization (excitation by light of alternating helicity), open circles – in the presence of nuclear polarization (helicity is constant).

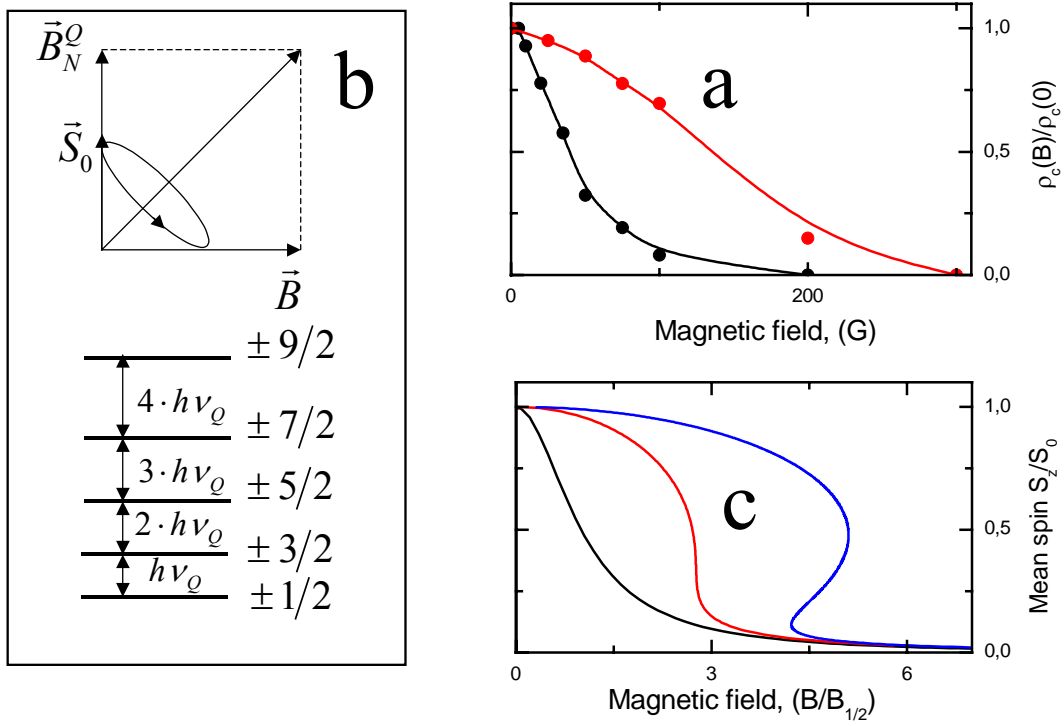

(a) Normalized data from Fig.1a. (b) Illustration of mutual orientation of zero-field mean spin $\vec{S}_0$, nuclear field $\vec{B}_N^Q$, external magnetic field $\vec{B}$ and energy level scheme of In in the presence of quadrupole interaction. (c) The result of calculation with the use of Eq.(2) under *q=0* (black curve); *q=5* (red curve); and *q=10* (blue curve).

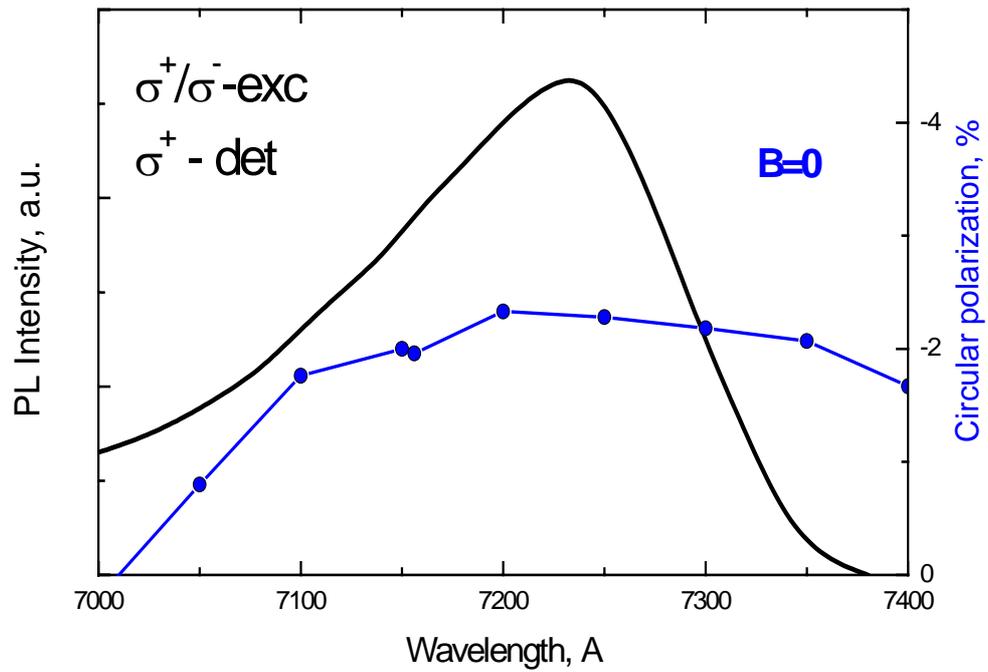

Photoluminescence spectrum (solid line) and circular polarization spectrum (blue points) of InP islands are measured in zero magnetic field in the absence of Overhauser effect.

REFERENCES

[1] I.A. Merkulov, Al.L. Efros, and M. Rosen, Physical Review B **65**, 205309 (2002)

[2] R. I. Dzhioev, V. L. Korenev, I.A.Merkulov, B. P. Zakharchenya, D. Gammon, Al.L. Efros, and D.S. Katzer, Physical Review Letters **88**, 256801 (2002); P.-F. Braun, X. Marie, L. Lombez, B. Urbaszek, T. Amand, P. Renucci, V. K. Kalevich, K.V. Kavokin, O. Krebs, P. Voisin, and Y. Masumoto, Physical Review Letters **94**, 116601 (2005)

[3] J.M. Elzerman, R. Hanson, L. H. Willems van Beveren, B. Witkamp, L.M.K. Vandersypen and L. P. Kouwenhoven, Nature (London) **430**, 431 (2004)

[4] M.I. Dyakonov and V.I. Perel, Chapter 2 in *Optical Orientation*, edited by F. Meier and B.P. Zakharchenya (North-Holland, New York, 1984).

[5] R.I. Dzhioev, B.P. Zakharchenya, V.L. Korenev, P.E. Pak, M.N. Tkachuk, D.A. Vinokurov, and I.S. Tarasov, JETP Letters **68**, 745 (1998)

[6] R. Oulton, A. Greilich, S.Yu. Verbin, R.V. Cherbunin, T. Auer, D.R. Yakovlev, M. Bayer, I.A. Merkulov, V. Stavarache, D. Reuter, and A. Wieck, Phys. Rev. Lett. **98**, 107401 (2007)

[7] C.W. Lai, P. Maletinsky, A. Badolato, and A. Imamoglu, Phys. Rev. Lett. **96**, 167403 (2006)

[8] I.A. Akimov, D.H. Feng, and F. Henneberger, Phys. Rev. Lett. **97**, 056602 (2006)

[9] I.A. Merkulov and V.G. Fleisher, Chapter 5 in *Optical Orientation*, edited by F. Meier and B.P. Zakharchenya (North-Holland, New York, 1984).

[10] R.I. Dzhioev, B.P. Zakharchenya, V.L. Korenev, P.E. Pak, D.A. Vinokurov, O.V. Kovalenkov, and I.S. Tarasov, Physics of the Solid State **40**, 1587 (1998)

[11] S. Cortez, O. Krebs, S. Laurent, M. Senes, X. Marie, P. Voisin, R. Ferreira, G. Bastard, J-M. Gérard, and T. Amand, Phys. Rev. Lett. **89**, 207401 (2002)

[12] A.S. Bracker, E. Stinaff, D. Gammon, M. E. Ware, J. G. Tischler, D. Park, A. Shabaev, Al.L. Efros, D. Gershoni, V. L. Korenev, I. A. Merkulov, Phys. Rev. Lett. **94**, 047402 (2005)

[13] Hole spin does not precess in low fields due to small transverse g-factor value. Spin precession of electrons in excitons is blocked by the electron-hole exchange: R.I. Dzhioev, V.L. Korenev, M.V. Lazarev, V.F. Sapega, D. Gammon, A.S. Bracker, Phys. Rev. B, **75** 033317 (2007)

[14] E.L. Ivchenko and G.E. Pikus. *Superlattices and other Heterostructures. Symmetry and Optical Phenomena*, v.110 (Springer-Verlag, 1995)

[15] A.A. Sirenko, T. Ruf, A. Kurtenback, and K. Eberl, in Proceedings of the 23rd International Conference "Physics of Semiconductors", Vol.2, (Berlin, 1996), p.1421

[16] Different contributions into the spin lifetime are possible: (i) typical island contains ~$2*10^6$ nuclei, giving the random nuclear field value about 45 G and dephasing time 1.5 ns; (ii) renewal of the resident electron due to the recombination (~1 ns) or tunneling processes; (iii) spin exchange with the non-polarized electrons of wetting layer.

[17] Similar Hanle curve broadening has been observed previously in bulk $HgI_2$ semiconductor. It was also explained by the effect of quadrupole interaction, see R.I. Dzhioev, and Yu.G. Kusraev Physics of the Solid State **38**, 1183 (1996)

[18] Quadrupole interaction is absent for the phosphor nuclei with *I=1/2*

[19] A. Löshe, Kerninduktion, veb Deutscher Verlag der Wissenschaften, Berlin, 1957.

[20] Ternary compound InGaP forms the barrier. In contrast to the As nuclei, the phosphorus nuclei do not possess quadrupole moment. In turn, the nearest neighbors of In and Ga (phosphorus nuclei) are not perturbed. Hence the QI in the barrier is absent, too.

[21] A.Kurtenbach, K. Eberl, and T. Shitara, Appl. Phys. Lett., **66** 361 (1995)

[22] C.P. Slichter, Principles of magnetic resonance (Springer-Verlag, New York, 1992), 3rd ed.

[23] R.K. Sundfors, R.K. Tsui, and C. Schwab, Phys. Rev. B, **13** 4504 (1976)

[24] Spin relaxation time of (x,y) nuclear spin components is determined by the z-components of local fields and can be much faster (see Ref.[9])

[25] Excepting for the case of the anticrossing of nuclear energy levels when the external field should be perpendicular to the z-axis to within 1 degree, see Ref.[9].

[26] R. I. Dzhioev, B. P. Zakharchenya, V. L. Korenev, and M. V. Lazarev, Physics of the Solid State, **41** 2014 (1999)

[27] A. Abraham, The principles of nuclear magnetism, Oxford: Clarendon Press, 1961